\documentclass[prd,nofootinbib,amsfonts]{revtex4}
\usepackage{graphicx,bm,amsmath,color}
\usepackage[latin1]{inputenc}
\usepackage{bbold}
\usepackage{wasysym}
\usepackage[T1]{fontenc}

\newcommand{\be}{\begin{equation}}
\newcommand{\ee}{\end{equation}}
\newcommand{\beq}{\begin{eqnarray}}
\newcommand{\eeq}{\end{eqnarray}}

\newcommand{\Mp}{M_\text{P}}
\newcommand{\rhow}{\rho_\text{w}}
\newcommand{\pf}{p_\text{F}}
\newcommand{\xf}{x_\text{F}}
\newcommand{\ch}{\mathcal{M}_\text{Ch}}

\begin{document}

\title{Testing energy non-additivity in white dwarfs}

\author{J.M. Carmona}
\affiliation{Departamento de F\'{\i}sica Te\'orica,
Universidad de Zaragoza, Zaragoza 50009, Spain}
\author{J.L. Cort\'es}
\affiliation{Departamento de F\'{\i}sica Te\'orica,
Universidad de Zaragoza, Zaragoza 50009, Spain}
\author{R. Gracia-Ruiz}
\email{jcarmona@unizar.es, cortes@unizar.es, rodrigo.gracia.ruiz@gmail.com}
\affiliation{Departamento de F\'{\i}sica Te\'orica,
Universidad de Zaragoza, Zaragoza 50009, Spain}
\author{N. Loret}
\email{niccolo@accatagliato.org}
\affiliation{Dipartimento di Matematica, Universit\`a di Roma ``La Sapienza", P.le A. Moro 2, 00185 Roma, Italy\\
and Perimeter Institute for Theoretical Physics, 31 Caroline Street North,
Waterloo, ON, N2L 2Y5 Canada.}

\begin{abstract}
We consider a particular effect which can be expected in scenarios of deviations from Special Relativity induced by Planckian physics: the loss of additivity in the total energy of a system of particles. We argue about the necessity to introduce a length scale to control the effects of non-additivity for macroscopic objects and consider white dwarfs as an appropriate laboratory to test this kind of new physics.
We study the sensitivity of the mass-radius relation of the Chandrasekhar model to these corrections by comparing the output of a simple phenomenological model to observational data of white dwarfs.
\end{abstract}

\maketitle

\section{Introduction}

The relativistic energy-momentum dispersion relation of particles, $E^2-\vec{p}^2=m^2$, is involved in every particle physics measurement, and, as so, it has been tested in countless occasions. This has been done with large precision however only in the relatively ``low-energy'' regime we are familiar with (energies at or below the Fermi scale, of $\mathcal{O}\sim 100\,$GeV). In recent times, astrophysics has also offered the opportunity to test this relation at much higher energies~\cite{astrophysics}. In fact, the loss of Lorentz invariance is a rather generic feature of quantum gravity developments~\cite{qugra}, so that modifications to this dispersion relation are expected at energies close to the Planck mass $\Mp$. However, the presence of amplifying mechanisms, such as threshold effects~\cite{grilloGZK} or galactic distances of propagation~\cite{PiranJacob}, as well as high-precision experiments~\cite{atomosfrios} makes possible to consider a quantum gravity phenomenology at energies much lower than $\Mp$~\cite{ACQuGra}.



In order to test this relation, it is quite common to assume corrections of order $1/\Lambda$ as the dominant effect if the energies involved are much smaller than the ultraviolet scale $\Lambda$ which controls the new physics beyond special relativity (SR). In a quantum gravity context, one would take $\Lambda=\Mp$, but the idea of a modified dispersion relation (MDR) is more general than that. A typical parametrization in which rotational invariance is maintained and the corrections can be expanded in powers of the momenta and the inverse of $\Lambda$ is
\be
E^2-\vec{p}^2+\frac{\alpha_1}{\Lambda}E^3+\frac{\alpha_2}{\Lambda}E\vec{p}^2=m^2\,,
\label{MDR}
\ee
where $\alpha_1$ and $\alpha_2$ are two coefficients of order 1 (that is, $\Lambda$ signals the energy scale at which the corrections are of order 1).

Of course, energy is not an observer-invariant quantity, so that when one is comparing the typical ``energy'' scale of an experiment with the scale $\Lambda$, one is either assuming a particular (laboratory) frame, or considering an invariant quantity (such as the invariant mass or the energy in the center of mass system in special relativity). In the case of a \emph{violation} of Lorentz invariance, the only possibility is to consider a particular, privileged, system of reference~\cite{kostel}. However, it is also possible to go beyond SR by considering a \emph{deformation} of Lorentz invariance. This is the case of doubly special relativity (DSR) models~\cite{DSRrefs}. The phenomenology of Lorentz invariance violation (LIV) and DSR models are in general quite different, since the existence of a relativistic principle in the latter forces to add to a MDR also a modification in the composition laws (the sum of the energies and momenta) of a multiparticle system. A general parametrization of modified composition laws (MCL) at order $1/\Lambda$ is
\begin{align}
E_1\oplus E_2 \,&=\, E_1 + E_2 + \frac{\beta_1}{\Lambda} E_1 E_2 + \frac{\beta_2}{\Lambda} \, \vec{p}_1\cdot\vec{p}_2 \label{E+}\\
\vec{p}_1\oplus \vec{p}_2 \,&=\, \vec{p}_1 + \vec{p}_2 + \frac{\gamma_1}{\Lambda} E_1 \vec{p}_2 + \frac{\gamma_2}{\Lambda} \, \vec{p}_1 E_2+\frac{\gamma_3}{\Lambda} \vec{p}_1\times \vec{p}_2 \,.
\label{MCL}
\end{align}
again with coefficients $(\beta_i, \gamma_i)$ of order 1.

The idea of MCL can also be considered in more general contexts than those of DSR models. It can be shown~\cite{locandrelpple} that MCL, that is, nonlinear corrections to additive energy-momentum composition laws, cause the locality property of interactions between particles to be lost for a general observer. This in fact agrees with the notion of ``relative locality'' that has recently been proposed in terms of a geometric interpretation of the departures from SR kinematics~\cite{relativelocality}. In particular, both a MDR and MCL may be present as signals of a LIV. What the existence of a relativity principle makes is to establish specific relations between the coefficients appearing in the MDR and the MCL. In fact, as it is shown in Ref.~\cite{Carmona:2012un} the $\beta$ and $\gamma$ coefficients of the MCL determine the $\alpha$ coefficients in the MDR in a relativistic theory beyond SR:
\be
\alpha_1 \,=\, - \beta_1 {\hskip 2cm} \alpha_2 \,=\, \gamma_1 + \gamma_2 - \beta_2\,.
\ee

Since the idea of MCL is very general, compulsory in DSR scenarios and possible in LIV scenarios, we want to address in this paper the physical implications of an energy non-additivity. Naively one would think that the best place to look for signals of an energy non-additivity would be systems of a large number of particles; however, for a sufficiently large system, the nonlinear terms cease to be small corrections contradicting, for example, the well-known physics of macroscopic objects. This difficulty is related with the general question of how the correction to SR in the microscopic domain affects macroscopic physics, which in the DSR context, where MCL are a necessary ingredient of the models, it is known as the ``soccer-ball problem''. This problem by itself deserves a general discussion of the different alternatives to solve it and a comparison of the required assumptions and their plausibility~\cite{soccerball}. We will just note here that if gravity or the structure of space-time is in the origin of this non-additivity, then it is natural to think that the key for these corrections to be large should be not a large number of particles, but a large number of particles which are close enough. That is, the relevant variable controlling the size of the non-additivity should be the density of the system. Incorporating this idea would immediately provide a possible solution to the soccer-ball problem, since the effects would be important not just for macroscopic objects, but for objects of high enough density.

In Sec. II of the paper we will make the previous idea explicit, by introducing the notion of a ``coherence length'' for the non-additivity corrections. Then we will explore the possible systems where these corrections could be tested, and conclude that white dwarfs are a natural candidate. In fact modifications in white dwarf physics owing to departures of Lorentz invariance have been considered previously~\cite{Camacho,Gregg,GiovanniWD}. We will review in Sec. III the main conclusions of these works and will remark their differences with the ideas here explored. In Sec. IV we will present a phenomenological model including non-additivity corrections for the energy of a white dwarf and will calculate how the mass-radius relation that can be extracted from the Chandrasekhar model gets modified. A comparison with experimental data to obtain bounds on the parameters of the model (essentially, the coherence length) will be done in Sec. V, showing the sensitivity of white dwarfs to this new physics. Finally, we will present our conclusions in Sec. VI.

\section{Energy non-additivity: coherence length}

Our aim is to explore possible observational consequences of new physics parameterized at a kinematical level by a modification (non-additivity) of the energy composition law. As explained in the Introduction, this non-additivity emerges naturally in specific scenarios related to quantum gravity effects, but we will just consider it as a phenomenological model for the new physics, without entering into the physical mechanisms at work. A departure from locality~\cite{locandrelpple} or a hidden effective interaction as a remnant of a quantum structure of space-time are examples of possible ideas related to the origin of an energy non-additivity.

Let us assume that the total energy of a two particle system is given by Eq.~(\ref{E+}), which contains nonlinear corrections to the simple addition of energies of order 1/$\Lambda$. $\Lambda$ is a high (ultraviolet) energy scale which may be as large as the Planck energy scale if the new physics we are considering is a remnant of quantum gravity. Since we will never have direct access to particles with energies of the order of the Planck energy scale, in order to have observable consequences of the non-additivity we have to rely on the amplification of the kinematic modification that happens in a many particle system, for which Eq.~(\ref{E+}) is generalized to (see Ref.~\cite{Carmona:2012un}):
\be
\sum_{\oplus}E_i\equiv \sum_i E_i + \sum_{i<j} \left(\frac{\beta_1}{\Lambda}E_i E_j + \frac{\beta_2}{\Lambda}\vec{p}_i\cdot \vec{p}_j\right).
\label{plusdef}
\ee
Let us think, for the sake of simplicity, of an object made of $N$ particles with the same energy $E$. Then the total energy $E_T$ will be deformed to\footnote{To illustrate the point we will use only the $\beta_1$-term of Eq.~(\ref{plusdef}).}
\be
E_T=NE+\frac{\beta_1}{\Lambda}\left(\frac{N^2-N}{2}\right)E^2=E_\Sigma + \frac{\beta_1}{\Lambda}\frac{N-1}{2N}\,E_\Sigma^2\,,
\ee
where the sum of the contributions $E_\Sigma=\sum_i E_i=NE$ could be, in principle, much larger than the characteristic energy scale of the deformation $\Lambda$. Of course this rough description of macroscopic manifestation of the deformed composition law for momenta (soccer-ball problem) is in contradiction with many observations, and should consequently be ruled out. In this paper we will consider a simple way of including a restriction on the non-additivity effects: we recognize this phenomenon to be a coherence process so that it should be confined within a certain \emph{coherence length}.

It is clear that two particles that are not correlated in any way, and far apart from each other, are really independent systems so that the total energy will be just the sum of their energies, with no nonlinear corrections. Therefore one can expect that the modification of the energy composition law will be limited to particles separated by a distance smaller than the coherence length $\ell_c$. We do not know the physical origin of this restriction on the non-additive corrections and how this length scale $\ell_c$ emerges. An interesting possibility is that the non trivial composition law for energies were a consequence of the superposition of different wave functions in a given quantum system; the coherence length would then define the range of this superposition. In any case, it seems clear that this length must be a microscopic scale, because it should provide a solution to the soccer-ball problem, but it may well depend on the properties of the system under consideration, such as the type of particles, whether or not they are entangled, or on their quantum mechanical properties, such as their de Broglie or Compton wavelengths. The origin of this coherence length is a very interesting problem that deserves future attention; however, in the present work we will just consider the consequences of this idea and, to do so, in the following we will explore the kind of systems which may be more sensitive to this new physics.

The total energy $E_T$ of a discrete system of particles will be calculated from Eq.~(\ref{plusdef}), which can be rewritten as
\be
E_T=\sum_i E_i\left( 1 + \frac{1}{2}\frac{\beta_1}{\Lambda}\sum\limits_{
	\begin{subarray}{c}
	 i,j \\ d(i,j)<\ell_c
	\end{subarray}
	}  E_j\right) + \frac{\beta_2}{2\Lambda}\sum\limits_{
	\begin{subarray}{c}
	 i,j \\ d(i,j)<\ell_c
	\end{subarray}
	} \vec{p}_i\cdot \vec{p}_j \,.
\label{discrete}
\ee
Looking at the parenthesis, it is clear that only systems such that the energy contained in a region of the size of the coherence length is of the order of the ultraviolet energy scale will be sensitive to the energy non-additivity. For a system of constant density $\rho$, the parameter controlling the deviation from usual additivity of the energy is
\be
\delta \sim V\left(\frac{\rho}{\Lambda}\right),
\ee
if $V<V_c\equiv \ell_c^3$, or
\be
\delta \sim V_c\left(\frac{\rho}{\Lambda}\right)
\label{delta}
\ee
if $V>V_c$. That is, to highlight the correction we need to consider systems of high density which are large enough. This leads to consider compact objects in nature, such as white dwarfs, neutron stars, or black holes, as the best windows for this new physics. One can then try to incorporate corrections to additivity in the energy composition law for any of these systems and compare with observations in order to extract bounds on the coherence-length scales for different systems of particles. In terms of the density of water $\rhow$ and the Planck mass $\Mp$, Eq.~(\ref{delta}) gives
\be
\delta \sim \left(\frac{\ell_c}{2\times 10^{-4}\,\text{m}}\right)^3 \left(\frac{\rho}{\rhow}\right) \left(\frac{\Mp}{\Lambda}\right).
\label{estimate}
\ee
This simple estimate tells us that the physics of a white dwarf (WD), with a typical density of order $\rho\sim 10^6\rhow$, could explore coherence lengths of the order of $10^{-6}$\,m when the new physics is associated to the Planck scale ($\Lambda=\Mp$). Since neutron stars have densities nine orders of magnitude larger than that of a WD, they have a larger sensitivity to the new physics parameterized by a non-additive modification of the energy composition law. In this work we will focus however on white dwarfs, since both their physics and their experimental observation are simpler than for the other families of compact objects. Our approach will be to identify the Planck energy scale as the ultraviolet energy scale associated to the new physics and then try to get bounds of a possible energy non-additivity in this specific physical system. As we will see, a WD is composed essentially by two different subsystems: a gas of ions and a degenerate Fermi gas of electrons. We will then extract bounds for the coherence-length scales associated to each subsystem.

\section{Planckian effects in white dwarfs}
\label{WDsection}

White dwarfs are compact objects which are produced from stars with masses similar to that of our Sun, $M_\odot=1.99\times 10^{30}$\,kg. Their radius are of the order of the Earth radius ($R_\oplus=6.3\times 10^6$\,m), and their characteristic density is of the order of $10^6$\,g/cm$^3$. A WD can be described as a gas composed by nuclei of light elements plus a completely degenerated electron gas. In a good approximation, the degenerate pressure of the electron gas compensates the gravitational attraction which is due to the mass of the ions. This equilibrium is no longer possible above a certain density of nuclei, which causes the collapse of the star for masses larger than around $1.4 M_\odot$. This maximum value of the mass of a WD (the Chandrasekhar mass) signals the theoretical limit which corresponds to the radius of the WD going to zero and infinite density of the star. However, this limit cannot be reached in practice, since neutronization processes start at densities around $10^7\,$g/cm$^3$, and for densities above $10^{12}\,$g/cm$^3$ the main contribution to the pressure is given by a newly formed gas of free neutrons, so that the star can no longer be considered as a WD.

Previously to this work, white dwarfs had already been chosen as laboratories to study possible deviations of Lorentz invariance. In Ref.~\cite{Camacho}, A. Camacho considered the consequences of a modification in the dispersion relation of particles similar to Eq.~(\ref{MDR}), specifically in the Chandrasekhar model~\cite{Chandra} giving the mass-radius relation of a WD. He obtained an amplification of Planck-scale effects in the limit in which the mass of the WD formally approaches the Chandrasekhar limit. In fact, it is easy to see why such a MDR cannot have any effect in the normal regime of white dwarfs: the maximum energy $E$ of the electrons in the Fermi gas is given by the Fermi momentum $\pf$, which is of the order of MeV for a typical WD. Since the corrections to the MDR are of order $E/\Mp$, they are completely negligible. The Chandrasekhar limit corresponds to the ultrarelativistic limit of a WD, $\pf\to\infty$, but in fact the amplification corresponding to this limit is irrelevant in practice, since much before reaching the densities that approach the Fermi energy to the Planck energy scale, dynamical instabilities occur as explained above and the analysis of the Planckian corrections is no longer valid, as it is remarked in Ref.~\cite{Gregg}.

In Ref.~\cite{GiovanniWD} the research of Refs.~\cite{Camacho,Gregg} was extended by adding a modification in the momentum-space integration measure as a way to consider the possible role of the geometry of momentum space that appears in a non-trivial way in contexts such as the relative locality framework~\cite{relativelocality}. It was seen, however, that the deformation of the measure of integration appears on the same level as the terms induced by the modification of the dispersion relation, and is therefore equally unobservable.

The effect of new physics that we are exploring in this paper, a modification in the energy composition law, is perhaps the third of the most popular ingredients (together with modifications in the dispersion relation and the role of the geometry of momentum space) that are present in the ``quantum-gravity folklore'' as low-energy Planckian effects. In this case, however, one expects that white dwarfs may give relevant bounds because of the amplification mechanism given by a large number of particles in a high density system as explained in the previous Section. We will now compute the modification to the mass-radius relation of the Chandrasekhar model which is produced by the non-additive corrections, and compare it afterwards with observational data in order to get the sensitivity of white dwarfs to this Planckian effect.

\section{Corrections from energy non-additivity to the Chandrasekhar model of white dwarfs}

The main features of the relationship between the mass of a white dwarf and its radius are given by the Chandrasekhar model~\cite{Chandra}, which assumes a steady state in which there is a hydrostatic equilibrium between the gravitational force and the pressure force at every point of the star. A non-additivity in the energy will modify this equilibrium condition by introducing corrections in the calculation both  of the mass (given essentially by the ion gas) and of the pressure (given by the degenerate electron gas). The sensitivity of white dwarf physics to a Planckian non-additivity will depend on the magnitude of the coherence lengths of these two subsystems. Our objective will be to compare these modifications of the mass-radius relation with experimental data in order to extract bounds to these parameters of our model.

In order to get a feeling of the corrections, we will first make a straightforward computation in a simple approximation to the problem, in which we assume a constant density for the complete star.

\subsection{Constant density approximation}

Let $U$ be the internal energy of the electron gas, and $E_G$ the gravitational energy of a uniform sphere of mass $M$ and radius $R$, $E_G=-3 G M^2/5R$, where $G$ is the gravitational constant. Then we can express the equilibrium condition with respect to the radius of the star as
\be
\frac{d (U+E_G)}{dR}=0 \quad \Rightarrow \quad P=\frac{3}{20}\frac{G M^2}{\pi R^4},
\label{eq}
\ee
where we introduced the pressure $P$ of the electron gas as $dU=-P\, dV=-P\, 4\pi R^2\,d R$.

In absence of non-additive corrections, the calculation of the internal energy $U$ is simply
\be
U=\frac{2}{h^3}\int \limits _{V}d\vec{r}\int\limits _{0}^{\pf}E(p)d\vec{p},
\label{usual U}
\ee
where $E(p)=\sqrt{p^2+m_e^2}$, and $m_e$ is the electron mass.
In the case of constant density, the Fermi momentum $\pf$ does not change with the distance to the center of the star, and can be written as
\be
\pf=\frac{h}{2\pi}\left(\frac{3\pi ^2 N_e}{V}\right)^{1/3},
\label{pf}
\ee
where $h$ is the Planck constant,\footnote{We will work in natural units, $c=\hbar=1$, but keep track of the $h$ factors, which, in these units, give factors of $(2\pi)$.} and $N_e$ is the number of electrons in the gas. This can be expressed in terms of the total mass of the star, as
\be
M=N_e\mu_e m_u,
\label{Mtotal}
\ee
where $m_u$ is the atomic mass unit, and $\mu_e$ is the mean atomic weight per electron. For a completely ionized element of mass number $A$ and atomic number $Z$, $\mu_e = A/Z$ to about 1 part in $10^4$~\cite{Shapiro}. The value of $\mu_e$ depends on the composition of the white dwarf; for example, for a carbon-oxygen white dwarf, $\mu_e=2$.

From Eqs.~(\ref{usual U}),~(\ref{pf}) and~(\ref{Mtotal}), it is easy to see that the pressure, calculated as $P=-d U/d V$, can be written as a function of $M$ and $R$ through the combination $M/R^3$. Then, Eq.~(\ref{eq}) gives an expression of the form
\be
M^2/R^4=f(M/R^3),
\label{MRR}
\ee
where $f$ is a certain function, defining a relation between the mass and radius of the star in the equilibrium. The expression of the function $f$ is simple in the non-relativistic and ultrarelativistic limits, defined as $\xf\equiv\pf/m_e \ll 1$ and $\xf \gg 1$, respectively. In the former case, taking $E(p)\approx p^2/2 m_e$ and inserting this into Eq.~(\ref{usual U}), one gets
\be
\frac{\mathcal{M}^2}{\mathcal{R}^4}=3.3\times 10^{-2}\,\mu_e^{-5/3}\,\tilde{\rho} ^{5/3},
\ee
where we have expressed the mass and radius in terms of their corresponding solar quantities, $\mathcal{M}\equiv M/M_\odot$ and $\mathcal{R}\equiv R/R_\odot$, and defined $\tilde\rho\equiv \mathcal{M}/\mathcal{R}^3$. Taking into account that $\mathcal{M}^2/\mathcal{R}^4=\mathcal{M}^{2/3}\tilde{\rho} ^{4/3}$, the previous expression can be rewritten as
\be
\mathcal{M}^{2/3} = 3.3\times 10^{-2}\,\mu_e^{-5/3}\,\tilde{\rho} ^{1/3},
\ee
and we get the well-known behavior $\mathcal{R}\sim \mathcal{M}^{-1/3}$. In the ultrarelativistic limit, $E(p)\approx p$, and we obtain
\be
\frac{\mathcal{M}^2}{\mathcal{R}^4}=3.68\,\mu_e^{-4/3}\,\tilde{\rho} ^{4/3},
\ee
so that
\be
\mathcal{M}^{2/3}=3.68\,\mu_e^{-4/3},
\label{usualUR}
\ee
or $\mathcal{M}=7.0\,\mu_e^{-2}$. This is the Chandrasekhar mass limit we mentioned in Sec. III; as we will see in the next subsection, its correct value (without the approximation of constant density), is $\ch=5.8\,\mu_e^{-2}$, which, for $\mu_e=2$, gives the typically quoted value of $\ch=1.45$.


\begin{figure}
  \centering
    \includegraphics[scale=0.3]{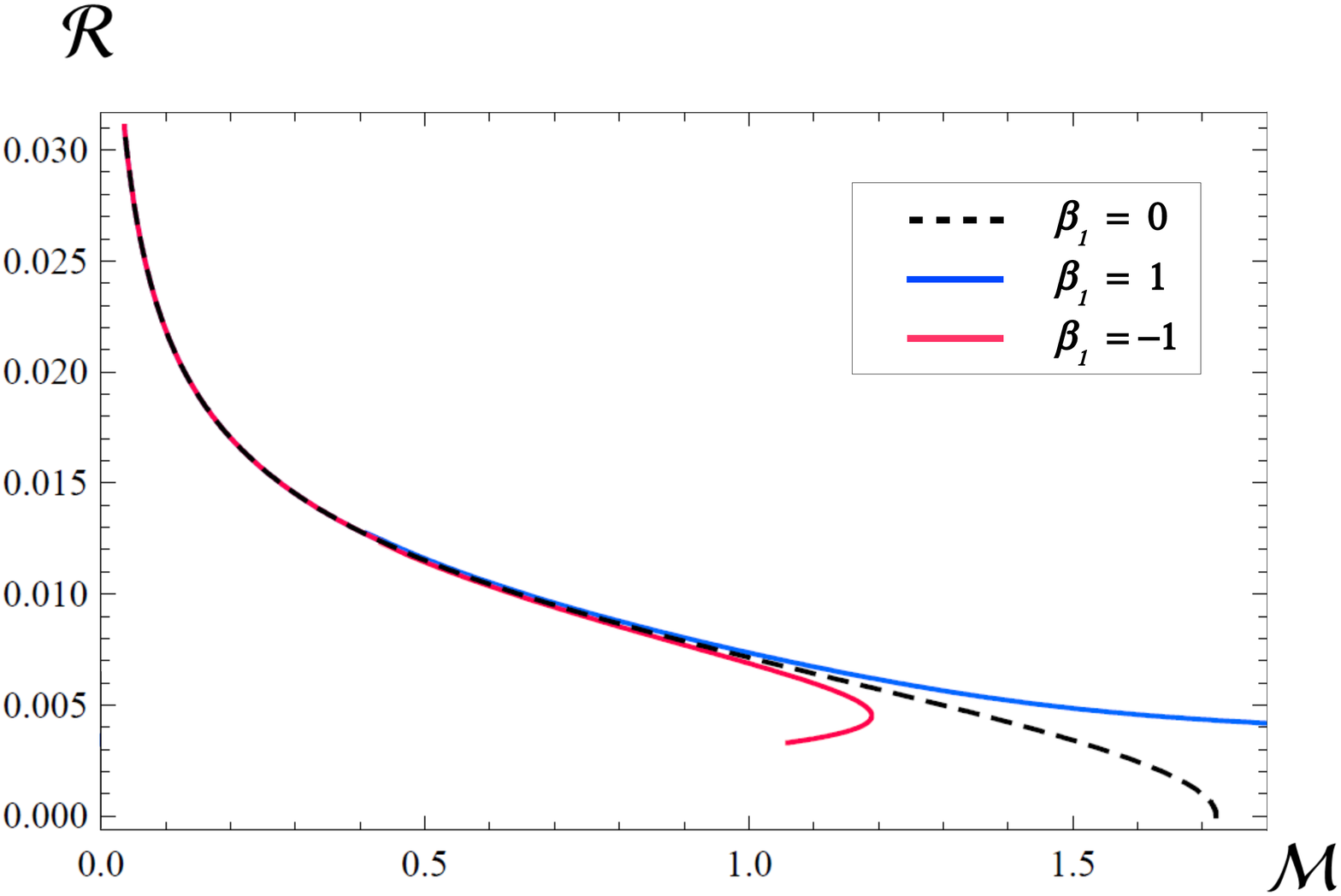}
  \caption{Mass-radius curve (in solar units of mass and radius) calculated in the constant density approximation, for $\mu_e=2$. The black dashed line represents the Chandrasekhar model, while the red and the blue ones show the correction on the behavior of the fermion gas due to MCL with a coherence length of three micrometers ($\ell_c = 3\times 10^{-6}$m) respectively for $\beta_1=-1$ and $\beta_1=1$.}
  \label{fig:densconst}
\end{figure}

The black dashed line of Fig.~(\ref{fig:densconst}) shows the characteristic curve for the relationship between mass and radius of white dwarfs, using the complete function $f$ in Eq.~(\ref{MRR}). Points on the right of this plot correspond to more compact stars (larger masses and lower radii), which are therefore denser, and for which the electrons of the Fermi gas are more relativistic, while the non-relativistic limit corresponds to the left side of the plot.

Now, the non-additivity in the energy produces two kinds of corrections. In the first place, the calculation of the internal energy $U$ in Eq.~(\ref{usual U}) has to be changed, which will produce a modification in the pressure of the electron Fermi gas. The new expression for $U$ is the continuum version of Eq.~(\ref{discrete}):
\begin{equation} \label{newU}
\begin{split}
U=\frac{2}{h^3}\int\limits _Vd\vec{r}\int\limits _0^{\pf} E(p)d\vec{p}& +\frac{\beta _1}{2\Mp}\frac{4}{h^6}\int\limits _Vd\vec{r}_1			 \int\limits _{V_c^{(e)}}d\vec{r}_2\int\limits _0^{\pf}  E(p_1)d\vec{p}_1\int\limits _0^{p_1}  E(p_2)d\vec{p}_2 \\
& +\frac{\beta _2}{2\Mp}			\frac{4}{h^6}\int\limits _Vd\vec{r}_1\int\limits _{V_c^{(e)}}d\vec{r}_2\int\limits _0^{\pf} \int\limits _0^{p_1}(\vec{p}_1\cdot\vec{p}_2) \, d\vec{p}_1	d\vec{p}_2\,,
\end{split}
\end{equation}
where we have limited some spatial integrals to the coherence volume of the electron gas, $V_c^{(e)}$, so that the nonlinear products of energy involve only spatial points inside this coherence volume. Symmetry considerations anticipate that the last two integrals in the $\beta_2$ term of Eq.~(\ref{newU}) are identically zero, so that the correction in $U$ will be proportional only to $\beta_1$, and in fact, it can be seen that it will be proportional to the product ($\beta_1 V_c^{(e)}$).

The second effect produced by the energy non-additivity is a correction in Eq.~(\ref{Mtotal}). In fact, the total mass of the star will be
\begin{equation}
\begin{split}
M&=\mu _e m_u \frac{2}{h^3}\int\limits _Vd\vec{r}\int\limits _0^{\pf}d\vec{p}+(\mu _e m_u)^2\frac{\beta _1}{2\Mp}\frac{4}{h^6}\int\limits _Vd\vec{r}_1			\int\limits _{V_c^{(i)}}d\vec{r}_2\int\limits _0^{\pf}  d\vec{p}_1\int\limits _0^{p_1} d\vec{p}_2 \\
&=(N_e\mu_e m_u)\left[1+\frac{3(\beta _1V_c^{(i)})(N_e\mu_e m_u)}{16\Mp\pi R^3}\right],
\end{split}
\label{newM}
\end{equation}
where $V_c^{i}$ now is the coherence volume associated to the gas of ions.

It is clear that both corrections will be larger for larger values of the Fermi momentum $\pf$ (for larger values of the energy of each individual electron), so that the larger deviations from the Chandrasekhar mass-radius curve will be observed in the ultrarelativistic limit. Treating both effects (the correction to the pressure of the Fermi gas, controlled by $V_c^{(e)}$, and the correction to the mass, controlled by $V_c^{(i)}$) separately, Eq.~(\ref{usualUR}) will be modified to
\be
\mathcal{M}^{2/3}=3.68\,\mu_e^{-4/3}\left[1+\alpha (\beta_1 \mathcal{V}_c)\mu_e^{-x}\tilde\rho^{y}\right],
\label{modifgeneral}
\ee
where $\mathcal{V}_c$ stands for the coherence volume ($V_c^{(e)}$ if we are considering the correction to the pressure, or $V_c^{(i)}$ if it is the correction to the mass) in Planck units (that is, $\mathcal{V}_c=V_c\cdot \Mp^3$),
$x$ is a numerical constant, and $\alpha$ is also a dimensionless quantity. A straightforward calculation gives
\begin{subequations} \label{numbers}
\begin{alignat}{2}
\alpha & =1.6\times 10^{-99} \quad  & x=y=4/3 & \quad \text{for the correction proportional to $V_c^{(e)}$} \label{alfa_e} \\
\alpha & =-9.0\times 10^{-95} \quad & x=0 \quad y=1 & \quad \text{for the correction proportional to $V_c^{(i)}$} \label{alfa_i}
\end{alignat}
\end{subequations}
The small numbers we get for $\alpha$ are just a reflection of the sensitivity of this system to a coherence length which is many orders of magnitude different from the Planck length [recall Eq.~(\ref{estimate}) and the fact that $\mathcal{V}_c$ in Eq.~(\ref{modifgeneral}) is expressed in Planck units], which is of course necessary to get an amplification for the non-additivity of the energy.

Note that the corrections of the electron gas and the ion gas in Eq.~(\ref{numbers}) are of different sign. Depending on the sign of $\beta_1$, one of the two will correct the Chandrasekhar mass of white dwarfs towards a lower value, while the other will extend the validity of the mass-radius relation to mass values larger than the standard Chandrasekhar limit. This can be observed in the red and blue lines of Fig.~(\ref{fig:densconst}), where it is shown the effect of the correction to the pressure of the eletron gas in the whole range of white dwarf masses for negative and positive $\beta_1$, respectively, and a value of the coherence length of the order of the micrometer (which, as commented after Eq.~(\ref{estimate}), is the expected value to produce relevant corrections in WD physics), in the constant density approximation.

We will now check the validity of these conclusions in a full calculation beyond the constant density approximation.

\subsection{Variable density: full calculation}

A more rigorous theoretical description of white dwarfs takes into account the variation of the density with respect to the center of the star. Our starting point will be the equilibrium equation
\be
\frac{dP}{dr} = - G \frac{m(r)}{r^2} \mu_e m_u n_e(r)
\label{eqec}
\ee
where we are assuming a spherically symmetric system with a radial coordinate $r$ and a density dominated by the rest mass of the ions, $P$ is the pressure of the degenerate electron gas, $n_e(r)$ the number density of electrons, and $m(r)$ the total mass contained in a sphere of radius $r$.

The pressure of the electron gas $P$ can be obtained from the electron energy density $\epsilon_e(r)$ through
\be
P = n_e^2\, \frac{\partial(\epsilon_e/n_e)}{\partial n_e} = - \epsilon_e + n_e\, \frac{\partial\epsilon_e}{\partial n_e} \,.
\ee
Taking into account that the number density of electrons in the degenerate electron gas is proportional to $\pf^3(r)$, where $\pf(r)$ is the Fermi momentum, one has
\be
\frac{dP}{dr} = - \left[\frac{2}{3} + \frac{1}{3} \frac{\pf\, d^2\pf/dr^2}{(d\pf/dr)^2}\right] \frac{d\epsilon_e}{dr} + \frac{1}{3} \frac{\pf}{d\pf/dr} \frac{d^2\epsilon_e}{dr^2} \,.
\label{dP}
\ee

Let us now calculate $\epsilon_e(r)$. The effect of the modification of the energy-momentum composition law on the energy density of the electron gas is given by
\be
\begin{split}
\epsilon_e(r) & = \frac{2}{h^3} \int\limits_0^{\pf(r)} dp_1 \,4\pi p_1^2\, (p_1^2+ m_e)^{1/2}  \\
&+ \frac{\beta_1  V_c^{(e)}}{2 \Mp} \frac{4}{h^6}  \int \limits_0^{\pf(r)} dp_1 \,4\pi p_1^2 \,(p_1^2+ m_e^2)^{1/2} \int\limits_0^{p_1}  dp_2\, 4\pi p_2^2\, (p_2^2 + m_e^2)^{1/2}
\end{split}
\ee
where, as before, $V_c^{(e)}$ is the coherence volume (volume of a sphere of radius the coherence length) of the electron gas system.

We can now use Eq.(\ref{dP}) to calculate the (radial derivative of the) pressure:
\be
\begin{split}
\frac{dP}{dr} &= \frac{4\pi}{3} \left(\frac{2}{h^3}\right) \frac{\pf^4}{(\pf^2 + m_e^2)^{1/2}} \, \frac{d\pf}{dr} \\
&+ \frac{\beta_1  V_c^{(e)}}{2 \Mp} \frac{4}{h^6} \, \frac{d\pf}{dr} \frac{4\pi}{3} \left[\frac{\pf^4}{(\pf^2+ m_e^2)^{1/2}} \int\limits_0^{\pf(r)} dp_1\ 4\pi p_1^2\, (p_1^2+ m_e^2)^{1/2} + 4\pi \pf^5\, (\pf^2+ m_e^2)\right] \,.
\end{split}
\ee
Introducing the electron Compton wave length $\lambda_e = 1/m_e$ and the dimensionless variable $\theta=(\pf^2/m_e^2 +1)^{1/2}$, which essentially gives us information about the density at a certain point, we have a simplified expression
\be
\frac{dP}{dr} = m_e n_e \frac{d\theta}{dr} \left[1 + b_e f_e(\theta)\right]\,
\label{dP/dr}
\ee
with
\be
b_e = \left(\frac{m_e}{\Mp}\right) \, \left(\frac{\beta_1 V_c^{(e)}}{\lambda_e^3}\right)
\label{be}
\ee
and
\be
f_e(\theta) = \frac{1}{16 \pi^2} \left[\theta (\theta^2 -1)^{3/2} + 9 \theta^3 (\theta^2 -1)^{1/2} - \ln\left(\theta + (\theta^2-1)^{1/2}\right)\right].
\label{fe}
\ee

The second effect of the non-additivity is the correction to the determination of the total mass $m(r)$ contained in a sphere of radius $r$. Assuming that the density is dominated by the rest-mass of the ions one has
\begin{equation}
m(r)=\mu _e m_u \left(\frac{32\pi ^2}{3h^3}\right)\int\limits _0^rdr_1\,r_1^2\,\pf^3(r_1)+(\mu _e m_u)^2\frac{\beta _1}{\Mp}\left(\frac{8\pi }{3h^3}\right)^24\pi V_c^{(i)}\int\limits _0^rdr_1\,r_1^2\,\pf^6(r_1)
\label{newmass2}
\end{equation}
where $V_c^{(i)}$ is the coherence volume of the ion system. The non-additive correction to the (radial derivative of the) mass $m(r)$ can be expressed (in a similar way as the correction to the pressure) as a multiplicative factor
\be
\frac{dm}{dr} = \frac{4}{3\pi} \mu_e m_u \frac{r^2}{\lambda_e^3} (\theta^2 -1)^{3/2} \left[1 + b_i f_i(\theta)\right]
\label{dm/dr}
\ee
where
\be
b_i = \left(\frac{\mu_e m_u}{\Mp}\right) \, \left(\frac{\beta_1V_c^{(i)}}{\lambda_e^3}\right)
\label{bi}
\ee
and
\be
f_i(\theta) = \frac{(\theta^2-1)^{3/2}}{18 \pi^2}\,.
\label{fi}
\ee
If we use the modified radial derivative of the pressure (\ref{dP/dr})  in the equilibrium equation (\ref{eqec}) we get
\be
m(r) = - \frac{\Mp^2}{\mu_e m_u} \frac{r^2}{\lambda_e} \frac{d\theta}{dr} \left[1 + b_e f_e(\theta)\right],
\label{mr}
\ee
which can be combined with the modified radial derivative of the mass Eq.~(\ref{dm/dr})
to find a differential equation for $\theta(r)$ whose solution determines the internal structure of the WD. By introducing a dimensionless radial variable $\xi$ through
\be
r^2 = \frac{3\pi}{4} \frac{\Mp^2}{\mu_e^2 m_u^2} \lambda_e^2 \,\xi^2,
\label{xidef}
\ee
the differential equation that determines the internal structure of the WD modified by the non-additive corrections is
\be
\frac{1}{\xi^2} \frac{d}{d\xi}\left(\xi^2 \frac{d\theta}{d\xi} \left[1 + b_e f_e(\theta)\right]\right) = - (\theta^2 -1)^{3/2} \left[1 + b_i f_i(\theta)\right].
\label{ecuacion}
\ee

The solution to Eq.~(\ref{ecuacion}) will give us the value of the Fermi momentum and the density as a function of the distance to the center of the star. As we will see in the next subsection, this solution will also allow us to determine the $\mathcal{M}(\mathcal{R})$ relation. It will then be possible to use a comparison with observational data of WD to put bounds on the possible values of the coherence volume determining the sensitivity of white-dwarfs to non-additive new physics. In this analysis one has to consider the limitations of the phenomenological model, which will restrict the possible values of the initial condition for Eq.(\ref{ecuacion}), $\theta_c\equiv \theta(\xi=0)$.

Since both $f_e(\theta)$ and $f_i(\theta)$ in Eqs.~(\ref{fe}) and~(\ref{fi}) are monotonically increasing functions, for fixed values of the parameters $b_e$ and $b_i$ the sensitivity to the new physics increases when one considers WD with a higher central density. But $|b_e f_e(\theta)|$, $|b_i f_i(\theta)|$ should be less than one, otherwise there is no justification to neglect the higher order non-additive corrections. Although one should require the non-additive correction to be a small perturbation, the previous requirement at the center of the star will be enough in order to have a small correction at the global level. Another limitation on the central density comes from the consistency of a model which is based on just a degenerate electron gas and an ion system, which requires densities below
the neutron drip point. These conditions will impose upper bounds to the possible values of $\theta_c$, which can be obtained in the ultrarelativistic approximation $(\theta\gg 1)$ for which
\begin{equation}
\theta _c\approx (\rho _c/\rho _0)^{1/3}\quad \quad f_e(\theta)\approx\frac{5\theta ^4}{8\pi ^2}\quad \quad f_e(\theta)\approx\frac{\theta ^3}{18\pi ^2}\,,
\end{equation}
where $\rho _0=\mu _em_u/3\pi ^2\lambda _e^3$. Using the estimate $\rho _*\approx 10^5\rho_0$ for the neutron drip point, then one has the density restrictions
\begin{equation}
\theta _c<10^{5/3}\quad \quad \theta _c<\left(\frac{8\pi ^2}{5|b_e|}\right)^{1/4}\quad \quad \theta _c<\left(\frac{18\pi ^2}{|b_i|}\right)^{1/3}\label{condiciones}.
\end{equation}
Note that the previous bounds on $\theta_c$ depend on the parameters of the non-additive corrections, $b_e$ and $b_i$. In order to compare with observations, in Sec.~\ref{results} we will consider appropriate values for $b_e$ and $b_i$ which allow to explore the catalogue of available data of white dwarfs.

\subsection{Resolution of the differential equation}

Eq.~(\ref{ecuacion}) is a second-order differential equation that can be solved numerically in a simple way if we rewrite it as a set of two first order equations by introducing an auxiliary function $\phi$ through
\be
\phi(\xi) = \xi^2 \frac{d\theta}{d\xi} \left[1 + b_e f_e(\theta)\right]
\ee
so that from Eq.~(\ref{ecuacion}) one has
\be
\frac{d\theta}{d\xi} =\frac{\phi(\xi)}{\xi^2 \left[1 + b_e f_e(\theta)\right]} \quad \quad \text{and} \quad \quad
\frac{d\phi}{d\xi} = - \xi^2 (\theta^2 -1)^{3/2} \left[1 + b_i f_i(\theta)\right].
\label{eqset}
\ee
In fact the auxiliary function $\phi$ can be related to the total mass $m(r)$ contained in a sphere of radius $r$. Using Eq.~(\ref{mr}) and the definition of the dimensionless variable $\xi$ [Eq.~(\ref{xidef})] one has
\begin{equation}
m(r)=-\frac{\sqrt{3\pi}}{2}\Mp\left(\frac{\Mp^2}{\mu _e^2m_u^2}\right)\phi (\xi)\,.
\label{m-phi}
\end{equation}

Let us first consider the conventional case with $b_e=b_i=0$
\begin{equation}
\frac{1}{\xi ^2}\frac{d}{d\xi}\left(\xi ^2\frac{d\theta}{d\xi}\right)=-(\theta ^2-1)^{3/2}\,.
\label{ecuacion0}
\end{equation}
This equation implies that $\xi ^2d\theta /d\xi$ is a monotonically decreasing function of $\xi$, so that this is also the case for $d\theta /d\xi$. Since the physically relevant initial condition corresponds to $d\theta /d\xi|_{\xi =0}=0$,\footnote{This condition is given by the spherical symmetry of the problem: since, near the center of the star, $m(r)\approx {4\pi\rho _cr^3/3}$, Eq.~(\ref{eqec}) tell us that $dP/dr|_{r=0}=0$, and Eq.~(\ref{dP/dr}) shows that this is also the case for $d\theta /d\xi$.} this implies that $d\theta/d\xi <0$. On the other hand, $\theta ^2(\xi)\geq 1$, so that $\theta$ will reach the minimum value $(\theta=1)$ for a finite value $\xi_*$ or approach asymptotically to it when $\xi\to\infty$.\footnote{The possibility that $\theta$ approaches asymptotically to a value greater than 1 can be excluded since the limit when $\xi\to\infty$ of the left hand side of the differential equation~(\ref{ecuacion}) is equal to zero while the right hand side vanishes only for $\theta=1$.} A solution where $\theta(\xi_*)=1$ corresponds to a WD with a finite radius
\begin{equation}
R=\frac{(3\pi )^{1/2}}{2}\frac{\Mp}{\mu _em_u}\lambda _e\,\xi_* \,,
\label{Rsol}
\end{equation}
and the total mass of the WD can be obtained from Eq.~(\ref{m-phi})
\begin{equation}
M=-\frac{\sqrt{3\pi}}{2}\Mp\left(\frac{\Mp^2}{\mu _e^2m_u^2}\right)\phi (\xi_*) \,.
\label{Msol}
\end{equation}
Since the non-additive corrections are introduced as a small effect on Eq.~(\ref{ecuacion0}), it natural to assume that this simple interpretation of the solution of the differential equation when $b_e=b_i=0$ will be also valid to the general case Eq.~(\ref{ecuacion}). Numerically, we will see that this is the case so that we will be able to use Eqs.~(\ref{Rsol}) and~(\ref{Msol}) to obtain the mass-radius relation of white dwarfs including non-additive Planckian effects.

\section{Results and comparison with experimental data}
\label{results}

We solved numerically Eqs.~(\ref{eqset}) by introducing a discretization and applying the classical (fourth-order) Runge-Kutta method~\cite{Runge-Kutta}. We discuss now the results for different values of the parameters $b_e$ and $b_i$.

\begin{figure}
  \hspace{-2cm}
    \includegraphics[scale=1]{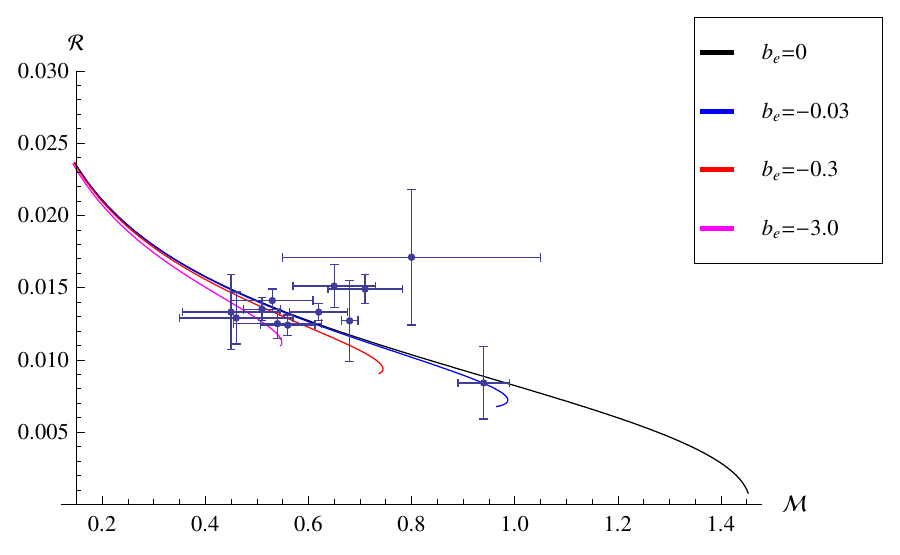}
    \includegraphics[scale=1]{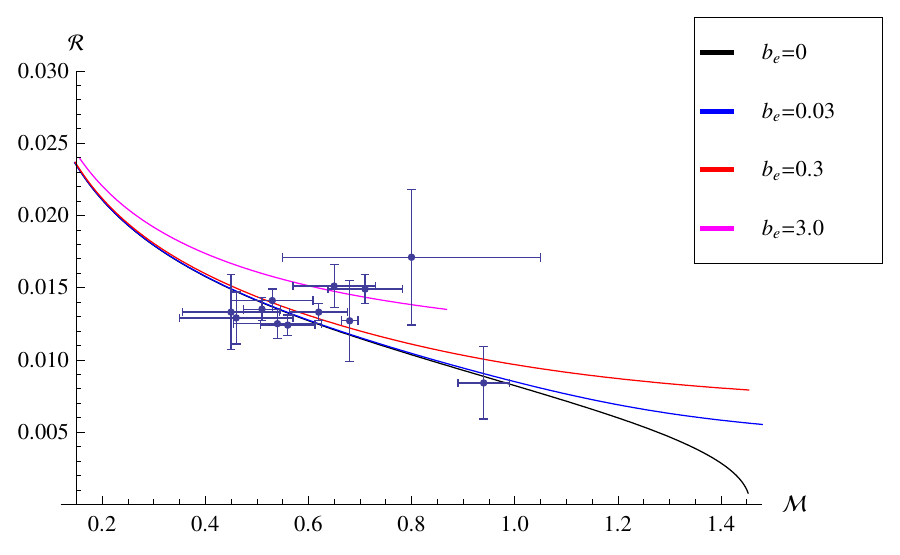}
  \caption{Mass-radius relation with a modified Fermi gas pressure for negative (left) and positive (right) values of $b_e$, for $\mu_e=2$, together with observational data of white dwarfs from Ref.~\cite{holberg}.}
  \label{fig:be}
\end{figure}

Fig.~\ref{fig:be} shows the mass-radius relation for different values of $b_e$, the parameter controlling the non-additive correction in the pressure of the electron Fermi gas, together with experimental points corresponding to a set of white dwarfs for which an exhaustive study of masses and radii has recently been carried out~\cite{holberg}. Each point in the curves correspond to a specific value of the central density ($\theta_c$).

If we compare the plots corresponding to $b_e>0$ (Fig.~\ref{fig:be}, right) with the Chandrasekhar model (black curve corresponding to $b_e=0$), we see that stars with a certain mass have larger radii and therefore are less dense. The curves end at the value of $\theta_c$ that limits the validity of the phenomenological model according to Eq.~(\ref{condiciones}), which decreases for larger corrections (larger values of $b_e$). The mass limit obtained for $b_e\to 0$ (the Chandrasekhar mass, $\mathcal{M}$=1.45), is clearly exceeded in this case, but the presence or absence of a larger mass limit will depend on details of the physics which go beyond the first order corrections that we analyze in this model. In any case, values of $b_e\sim\mathcal{O}(1)$ would be excluded by the experimental data (specifically, by the more relativistic white dwarf, the point more to the right of the plot, which corresponds to Sirius-B).

The curves corresponding to $b_e<0$ (Fig.~\ref{fig:be}, left) show the opposite behavior to the previous case. The main effect of the correction is to give smaller radii and larger central densities compared with Chandrasekhar models of the same mass, or, equivalently, for a given central density, the mass of the modified model is less than in the usual model. It is interesting to note that this type of behavior is very similar to that produced by standard electrostatic corrections to the Chandrasekhar model~\cite{hamada}. The magnitude of the effect in the present case, however, is controlled by the value of $b_e$ (by the coherence length defining the non-additivity in the electron gas). Fig.~\ref{fig:be} (left) shows that the upper mass limit of white dwarfs decreases when the correction gets larger. In particular, data from the more relativistic white dwarfs exclude values of $|b_e|$ larger than $\sim\mathcal{O}(10^{-1})$ for the negative case. From the definition $b_e$, Eq.~(\ref{be}), we can extract the bound
\begin{equation}
|\beta_1|\, V_c^{(e)}<10^{-16}\ \text{m}^3,
\end{equation}
that for a value of $\beta_1$ of order one, corresponds to a coherence length in the gas of electrons of the order of the $\mu$m.

\begin{figure}
\hspace{-2cm}
    \includegraphics[scale=1]{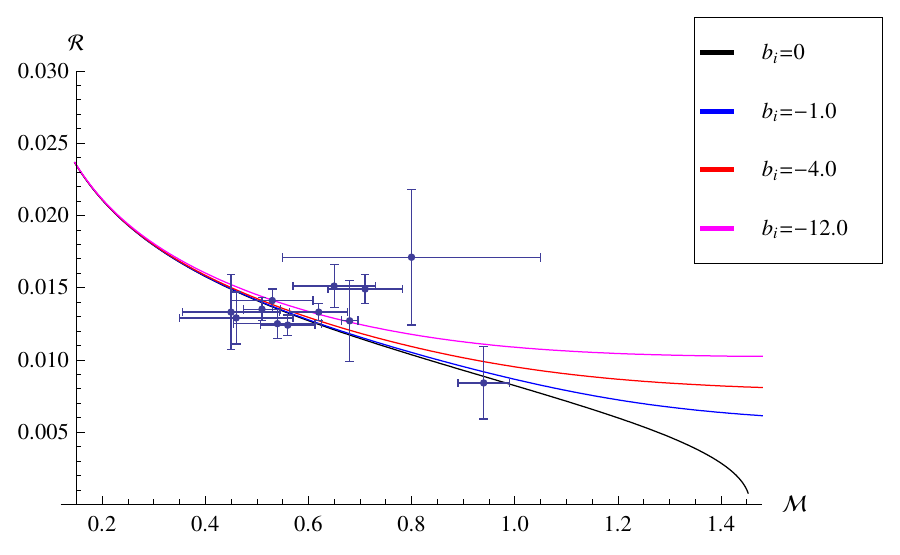}
    \includegraphics[scale=1]{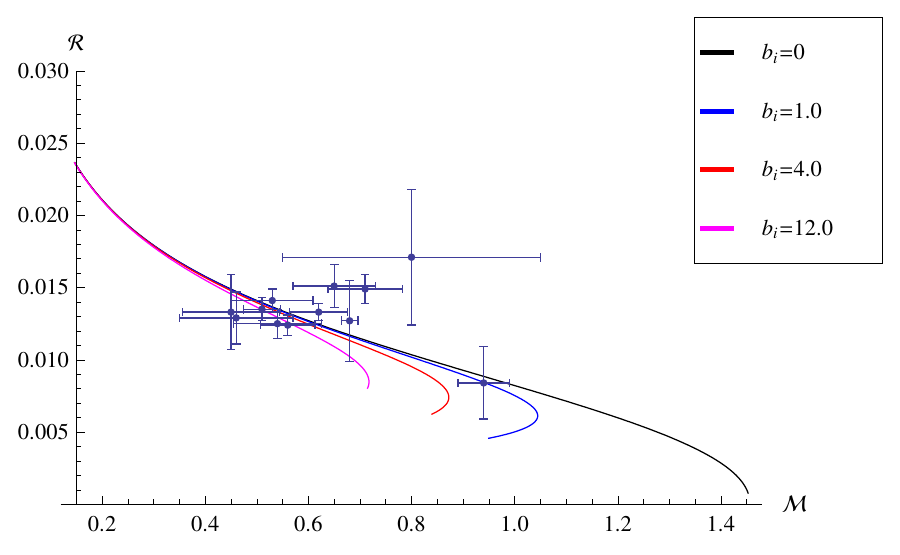}
  \caption{Mass-radius relation with a modification in the mass for negative (left) and positive (right) values of $b_i$, for $\mu_e=2$, together with observational data of white dwarfs from Ref.~\cite{holberg}.}
  \label{fig:bi}
\end{figure}

The curves corresponding to the correction in the mass of the WD are shown in Fig.~\ref{fig:bi}. As anticipated in the qualitative analysis with the constant density approximation, the behavior of this correction with respect to the sign of the parameter $\beta_1$ is the opposite to that in the pressure of the electron Fermi gas. As in the previous case, the comparison with experimental data can be used to extract bounds on $b_i$. Specifically, for the $b_i<0$ case, the bound is $|b_i|<\mathcal{O}(1)$, corresponding to
\begin{equation}
|\beta_1|\, V_c^{(i)}<10^{-19}\ \text{m}^3,
\end{equation}
which for a value of $\beta_1$ of order one, corresponds to a coherence length in the gas of ions of the order of tenths of $\mu$m. This shows that the correction to the mass of the ions is more sensitive (an order of magnitude in the coherence length, but three orders of magnitude in the coherence volume, to which the correction is proportional) that the correction to the pressure of the electron gas with respect to a non-additivity in the energy [note that the fact that the bound on $|b_i|$ is softer than the bound on $|b_e|$ is due to the different factors in the definitions of $b_e$ and $b_i$, Eqs.~(\ref{be}) and~(\ref{bi})].

\section{Summary and conclusions}

In this work we have explored from a phenomenological point of view a non-additive composition law for the total energy of a system of particles as a possible remnant of quantum gravity effects associated to a loss of Lorentz invariance. Since the effects are suppressed by the Planck scale, it is necessary to consider systems with a large number of particles in order to have an appropriate amplification of the corrections. However, one has to assume that the non-additivity is maintained only within distances lower than a certain coherence length in order to account for the energy of everyday macroscopic objects. The introduction of this scale makes the effect proportional to the density of the system, therefore restricting our attention to compact objects present in the Universe, such as white dwarfs, neutron stars or black holes. We chose to use white dwarfs as a laboratory for the new effects because their physics are well known and particularly simple, and the observational data can be used to put bounds on them. However, these bounds will be relevant only as long as the coherence length depends on the particular physical system. This seems reasonable if the origin of this length lies in the quantum properties of the system, as we argued in Sec. II. In the case of a universal coherence length for the non-additive effects, neutron stars would give much more restrictive bounds than white dwarfs because of their much greater density.

We considered the modifications in the Chandrasekhar model giving the mass-radius relation of white dwarfs produced by the non-additivity in the energy. They imply corrections both in the calculation of the pressure of the electron Fermi gas and in the relation between the total mass of the star and the masses of its constituents, which involve essentially the ion gas. A numerical analysis of the modified differential equation which results from the hydrostatic equilibrium condition and the comparison with experimental data has allowed us to obtain bounds on the coherence length of both the electron and the ion gas. These bounds reflect the sensitivity of white dwarf physics to a non-additivity in the energy.

We note that in order to obtain these phenomenological limits to the parameters of the model we had to consider separately the effects on the electron and on the ion gases. Since, as we remarked in the text, the two effects produce opposite corrections, it would be possible to get a cancellation in some range of values of the two parameters $b_e$ and $b_i$. However, the present accuracy of white dwarf observations only allow us to put bounds on the order of magnitude of these paramaters, and, to do so, it was a necessary simplification to assume that one of the effects dominates over the other. Future determinations of masses and radii of white dwarfs with greater accuracy, especially for high density stars, would allow us to put further restrictions to the model, in particular on the sign of $\beta_1$, the coefficient that gives the leading order Planckian correction to the energy composition law. In this sense, this work provides a new motivation to improve the observations of WD, in particular for those stars with a mass close to the Chandrasekhar limit.

As possible future research, it would be interesting to extend the present analysis to the other two types of very compact objects (neutron stars and black holes). Another possible manifestation of the new physics explored in this work is in the evolution of the primitive universe, including the inflationary epoch. It would also be of course very important to try to deepen in the notion of the coherence length and its interpretation along the lines we mentioned in Sec. II, which could offer a guide in the study of the implications of the energy non-additivity. Finally, even without a proper understanding of its origin, it would be interesting to work out in detail how the idea of the coherence length may provide a solution to the soccer-ball problem, which is a common ingredient in all extensions of SR.

As a final comment, let us recall (see the Introduction) that there are scenarios motivated by quantum gravity considerations that explore deviations of Special Relativity both at the level of a modified dispersion relation (MDR) and a modified composition law (MCL) in the energy or momentum of a system of particles, such as in DSR models. We argued in Sec.~\ref{WDsection} that a (Planck-scale) MDR cannot have any relevant effect on white dwarf physics, since the energies of the individual particles are much lower than the Planck energy scale. We have seen however that this is not the case with a MCL, owing to the amplification coming from the large number of particles that constitute the system. This is a peculiar situation from the point of view of DSR, in which the existence of a relativity principle produces a suppression of the Planckian effects owing to the constraints imposed by the consistency between the conservation laws of the energy-momentum, the dispersion relation, and the relativity principle~\cite{asr}. This is a general conclusion in the kinematics of a few (high-energy) particles, for which the effects of the MDR and the MCL are of the same order. The present study shows that the kinematics of a large system of particles is a counterexample of the naive belief that this cancellation should be generically present in relativistic theories beyond SR.

\section*{Acknowledgments}
This work is supported by the Spanish MINECO (grants FPA2009-09638 and FPA2012-35453) and DGIID-DGA (grant 2011-E24/2) and a ``La Sapienza'' fellowship (perfezionamento all'estero, area CUN1). It was also made possible in part through the support of a grant from the John Templeton Foundation. The opinions expressed in this publication are those of the author and do not necessarily reflect the views of the John Templeton Foundation. Research at Perimeter Institute is supported by the Government of Canada through Industry Canada and by the Province of Ontario through the Ministry of Economic Development and Innovation.

We acknowledge useful conversations with A. Grillo, and thank G. Amelino-Camelia for the organization of the workshop ``Planck scale in astrophysics and cosmology'' (Rome, March 2013), where we had the opportunity to present and discuss this work in its early stages.


\end{document}